\documentclass[english,conference]{IEEEtran}

\usepackage[T1]{fontenc}
\usepackage{xcolor}
\usepackage{amsmath}
\usepackage{amsthm}
\usepackage{amsfonts}
\usepackage{graphicx}
\usepackage{mathtools}
\usepackage{amssymb}
\usepackage{dblfloatfix}

\makeatletter
\theoremstyle{plain}
\newtheorem{thm}{\protect\theoremname}
\theoremstyle{plain}

\theoremstyle{plain}

\theoremstyle{definition}

\usepackage[caption=false,font=footnotesize]{subfig}

\makeatother

\providecommand{\corollaryname}{Corollary}
\providecommand{\lemmaname}{Lemma}
\providecommand{\theoremname}{Theorem}
\providecommand{\examplename}{Example}
\providecommand{\propositionname}{Proposition}
\providecommand{\claimname}{Claim}
\providecommand{\conjecturename}{Conjecture}
\providecommand{\definitionname}{Definition}

\DeclareMathOperator{\Tr}{Tr}

\usepackage[disable,textsize=tiny]{todonotes}

\begin{document}
\title{Faithful Simulation of Broadcast Measurements}
\author{\IEEEauthorblockN{Anders H{\o}st-Madsen\\}\IEEEauthorblockA{Department of Electrical \& Computer Engineering\\University of Hawaii, Manoa\\
Honolulu, HI, 96822, Email: ahm@hawaii.edu}}
\maketitle

\begin{abstract}
In this paper a central server Charlie has
access to a quantum system C and measures it
with a POVM \(\{\Lambda_x\}\). Alice and Bob are
only interested in the partial results \(g_A(x)\) respectively
\(g_B(x)\). Alice, Bob, and Charlie share common randomness
and Alice and Bob only need to faithfully simulate their
measurements. The paper develops to achievable regions for
the amount of communication needed to Alice and Bob.
\end{abstract}


\section{Introduction}
The aim of network information theory \cite{ElGamalKimBook}
is for nodes to obtain certain desired information generated
by sources. In some cases, this is the raw information generated
by the sources, but in many cases it is a function of the information.
For example, a node might only need the sum or average of certain measurements or need to make a decision based on the data.
The required rates for
computing functions in networks are therefore a widely
studied problem in classical information theory, for example
\cite{OrlitskyRoche01, DoshiMedardEffrosAl10,FeiziMedard14,GuangYeungAl19,Malak22, Gaspar08,NazerGastpar08,NazerGastpar07, SoundararajanVishwanathAl12, AppuswamyZegerAl11, LiAl23, LiMaddah-AliAl17, ZhuGunduz21, BergerZhang94, ViswanathanBerger97, HeAl16, Bar-Yossef11,ArbabjolfaeiKim18,ElRouayhebSprintsonAl10,EffrosAl15, SalehiAl25}.

Measurements are, of course, central to quantum theory.
A version of the networked function computation problem for
quantum measurements is as follows. 
Nodes $A,B,C\ldots$ have a multipartite system represented
by a density operator $\rho^{ABC\ldots}$. One node, say A,
wants to perform a global measurement represented
by POVM $\{\Lambda^{ABC\ldots}_a\}_a$, but has only its
local quantum system or no access to a quantum system at all. The measurement must be executed using local quantum instruments at each node, with the outcomes transmitted to a destination node. How much transmission is needed
to find $\{\Lambda^{ABC\ldots}_a\}_a$, which is, of course, classical information? An application could be a distributed quantum computer,
where one would like to extract a classical result that depends on
the total quantum system.

Winter \cite{Winter2004} introduced the idea of
faithful simulation of measurements. Suppose that
in a two node system
Alice performs some measurements and want to
transmit the result to Bob.
Alice performs a measurement on a quantum state $\rho$
and sends some classical bits to Bob, who intends
to \emph{faithfully} recover Alice's measurement,
preserving correlation with the reference system.
The key observation is that if Alice and Bob have
common randomness, the number of bits transmitted
from Alice to Bob can be decreased below that of
classical data compression of Alice's measurement 
outcomes, 
while still preserving the correlation with
the reference system. We refer the reader
to the overview paper \cite{WildeAl2012} and the
paper \cite{AtifHeidariSandeep22} for more details
about faithful simulation and applications of it. 

Networked measurements can be combined with faithful simulation. The paper \cite{AtifHeidariSandeep22}
 considered the following problem. Alice and Bob have a shared
bipartite quantum system $\rho^{AB}$. Alice makes 
measurements with a POVM $\{\Lambda^A_u\}$ on $\rho^A$ and
Bob measures with $\{\Lambda^B_v\}$ on $\rho^B$. Alice, Bob, and Charlie
share  common randomness, and Alice and Bob transmit some classical
information to Charlie. The goal is  for Charlie to
faithfully simulate a function $z=g(u,v)$ of the measurements.
In the paper \cite{Host26} the author strengthened this result
in the case when Alice communicates to Bob whob calculates
$z=g(u,v)$. 

In this paper we consider the opposite problem.
Charlie has access to a quantum system $C$. Alice would
like to get the results of measurements $\{\Lambda_{x_A}\}$ and
Bob $\{\Lambda_{x_B}\}$. In general $\{\Lambda_{x_A}\}$ and
$\{\Lambda_{x_B}\}$ might be incompatible. So, we consider
the following special case.
Charlie performs some measurements $\{\Lambda_x\}$ and Alice and Bob wants
functions $g_A$ and $g_B$ of the measurement results $x$. There are error
free links to Alice and Bob. This is somewhat related
to index coding for classical communications \cite{Bar-Yossef11,ArbabjolfaeiKim18,ElRouayhebSprintsonAl10,EffrosAl15, SalehiAl25}; however, we do not consider side-information
at the receivers in this paper, so in classical communications
the problem is trivial. However, if Alice and Bob only need
to faithfully simulate Charlie's measurement, it is possible to
reduce the communication rates.

The paper largely follows the notation and terminology established
in \cite{WildeBook}. We will use theorems and lemmas of \cite{WildeBook}
without repeating them here. 
We use implicit notation for probabilities when the meaning is unambiguous, e.g.,
$p(u^n)=p^n_U(u^n)$.
A sub-POVM is a set of operators $\{\Lambda_x\}_x$ with
$\Lambda_x\geq 0$ and $\sum_x \Lambda_x\leq I$.
For an integer $a$ $[a]=\{1,2,\ldots a\}$. We use $\|\cdot\|_1$ 
to denote trance norm and trace distance.
\section{Problem statement}
Charlie measures $\{\Lambda_x\}$ on a system C
with density matrix $\rho$. Alice is interested in
$x_A=g_A(x)$ and Bob in $x_B=g_B(x)$. Charlie can
of course calculate the functions and transmit
$x_A$ and $x_B$ directly, requiring $R_A=H(X_A)$ respectively
$R_B=H(X_B)$ bits. The question this paper seeks to answer is if this can be reduced
if only faithful simulation of the measurements are required.

A more precise statement is that Alice and Bob want
the measurement results
\begin{align}
  \Lambda_{x_A} &= \sum_{x: g_A(x)=x_A}\Lambda_x \\
  \Lambda_{x_B} &= \sum_{x: g_B(x)=x_B}\Lambda_x \label{eq:x_B}
\end{align}

Rather than directly measuring $\{\Lambda_x\}$, Charlie
can do a sequential measurement: first
Charlie measures with $\{\Lambda_{x_A}\}$ and transmits
the result to Alice. He then measures with $\{\Lambda_{x_B}\}$
on the system $\rho_{x_A}$ remaining after the first
measurement. The details are as follows. After measuring $\Lambda_{x_A}$ the  state of the system is
\begin{align*}
  \rho_{x_A}&=\frac 1{\Tr\{\Lambda_{x_A}\rho\}}\sqrt{\Lambda_{x_A}}\rho\sqrt{\Lambda_{x_A}}^\dagger
\end{align*}
Charlie than uses the following conditional measurements
\begin{align}
  \Lambda_{x_B|x_A}&= (\sqrt{\Lambda_{x_A}}^\dagger)^{-1}\sum_{x: g_A(x)=x_A, g_B(x)=x_B}\Lambda_x(\sqrt{\Lambda_{x_A}})^{-1}
  \label{eq:x_Bx_A}
\end{align}
Here the inverse is only in the subspace spanned by $\rho_{x_A}$. 
We notice that
\begin{align*}
  \sum_{x_B}\Lambda_{x_B|x_A} &=(\sqrt{\Lambda_{x_A}}^\dagger)^{-1}\sum_{x: g_A(x)=x_A}\Lambda_x(\sqrt{\Lambda_{x_A}})^{-1}
  = \Pi_{x_A}
\end{align*}
so that it is a valid POVM. Furthermore, because
$\forall x: g_A(x)=x_A, g_B(x)=x_B:\text{span}\{\Lambda_x\}\subset
\text{span}\{\Lambda_{x_A}\}$, so that the inverse is
in the correct subspace, measurement with  $\Lambda_{x_B|x_A}$
is equivalent with measuring 
\begin{align*}
  \Lambda_{x_A,x_B} &= \sum_{x: g_A(x)=x_A, g_B(x)=x_B}\Lambda_x
\end{align*}
on the original system $\rho$.

It should be noticed that (\ref{eq:x_B}) is not
the same \emph{measurement} as (\ref{eq:x_Bx_A}). In particular,
with (\ref{eq:x_Bx_A}) Bob is only informed of the result
$x_B$; he does not know $x_A$ and therefore does not know exactly
which measurement was made. We consider two measurements to be
equivalent if the statistics of the measurement results are identical.
This can be made precise with the ideas from faithful simulation.
Namely, define \cite{WildeAl2012}
\begin{align*}
  \mathcal{M}_{\Lambda_x}(\rho)=\sum_x \Tr\{\Lambda_x\rho\}|x\rangle\langle x|
\end{align*}
Two measurements $\{\Lambda_x\}$ and $\{\tilde \Lambda_x\}$ are
equivalent if
\begin{align}
  I^R\otimes \mathcal{M}_{\Lambda_x}(\phi_\rho)=I^R\otimes \mathcal{M}_{\tilde\Lambda_x}(\phi_\rho) \label{eq:equivalence}
\end{align}
where $\phi_\rho$ is a purification of $\rho$ and $I^R$ is the identity
map on the reference system. It is now easy to see that 
(\ref{eq:x_B}) and (\ref{eq:x_Bx_A}) are equivalent measurements.

\section{Faithful simulation}
In faithful simulation, the criterion (\ref{eq:equivalence}) is
weakened as follows. We say that $\{\tilde \Lambda_{x_A^n}\}$ faithfully
simulates $\{\Lambda_{x_A^n}\}$ on $\rho^{\otimes n}$ if for any $\epsilon>0$ there exists an $N$ so
that for all $n\geq N$
\begin{align*}
  \MoveEqLeft\left\|(I^R)^{\otimes n}\otimes \mathcal{M}_{\tilde \Lambda_{x_A^n}}(\phi_\rho^{\otimes n})-(I^R)^{\otimes n}\otimes \mathcal{M}_{\Lambda_{x_A^n}}(\phi_\rho^{\otimes n})\right\|_1 \nonumber\\
  &= \sum_{x_A^n}\left\|\sqrt{\rho^{\otimes n}}(\tilde\Lambda_{x_A^n}-\Lambda_{x_A^n})\sqrt{\rho^{\otimes n}}\right\|_1
  \leq\epsilon,
\end{align*}
where the equality is due to \cite[Lemma 4]{WildeAl2012},
and with an equivalent definition for $\{\Lambda_{x_B^n}\}$.

Let $R_A$ and $R_B$ be the rate of transmission to Alice/Bob and
$S_A$ and $S_B$ be the rate of common randomness. That is Charlie
and Alice share common randomness at a rate $S_A$ and Charlie
and Bob share common randomness at a rate $S_B$ (not shared with Alice). Additionally, Bob may have to know the common
randomness Charlie shares with Alice.
\begin{thm}
With sequential measurement (Alice before Bob), the
rate region for Alice is
\begin{align*}
  R_A &\geq I(X_A;R) \\
  R_A+S_A &\geq H(X_A)
\end{align*}
Bob has two possible rate regions. First,
\begin{align*}
  R_B &\geq I(X_A,X_B;R) \nonumber\\
  R_B+S_B &\geq H(X_B|X_A)
\end{align*}
Bob also needs to know Alice's common randomness with Charlie.
Second,
\begin{align*}
  R_B &\geq I(X_B;R,X_A) \nonumber\\
  R_B+S_B &\geq H(X_B)
\end{align*}
Bob does not need to know Alice's and Charlie's shared randomness. Here $R$ is a reference system purifying $\rho$.
\end{thm}
We do not have a tight converse for this problem. However,
the rate region is optimum in two trivial cases: if $g_A=g_B$,
the first region is optimum, and if $X_A$ and $X_B$ are independent,
the second region is optimum.

We will now outline how the faithful simulation measurement
operators are defined. It is very similar to \cite{WildeAl2012, Host26}, but modified for sequential measurement.

The Alice measurement is completely
standard resulting in the rate constraint $R_A\geq I(X_A;R)$
and $R_A+S_A\geq H(X_A)$ \cite{WildeAl2012}. We need a few results from
the proof of this result. Let
\begin{align}
  \tilde p(x_A^n)=\begin{cases}
  	\frac 1 S p(x_A^n) & w^n\in T_\delta^{X_A^n} \\
  	0& \text{otherwise}
  \end{cases} \label{eq:pwpruned}
\end{align}
with $S=\sum_{x^n\in T_\delta^{X_A^n}}p(x^n)$. Further, let
$\hat p(x_A^n)$ be the empirical distribution for the approximate
measurements of $x_A^n$. Then
\begin{align}
  \sum_{x^n_A\in T_\delta^{x_A^n}}
  |\tilde p(x_A^n)-\hat p(x_A^n)|\leq \epsilon \label{eq:pA}
\end{align}
with high probability (converging to zero as $n\to\infty$). This follows from
\cite[(30)]{WildeAl2012}. We also have
\begin{align}
  \sum_{x^n_A\in T_\delta^{x_A^n}}
  \left\|\tilde p(x_A^n)\rho_{x_A^n}-\hat p(x_A^n)\tilde\rho_{x_A^n}\right\|_1\leq \epsilon \label{eq:rxA}
\end{align}
where $\tilde\rho_{x_A^n}$ is the state after the approximate
Alice measurements. From (\ref{eq:pA}) and (\ref{eq:rxA})
we also get
\begin{align}
  \sum_{x^n_A\in T_\delta^{x_A^n}}
  \tilde p(x_A^n)\left\|\rho_{x_A^n}-\tilde\rho_{x_A^n}\right\|_1\leq 2\epsilon \label{eq:rxA2}
\end{align}

For the Bob measurement, we define
\begin{align}
  \hat \rho^A_{x_B|x_A} &=\frac 1{\Tr\{\Lambda_{x_B|x_A}\rho_{x_A}\}}\sqrt{\rho_{x_A}}\Lambda_{x_B|x_A} \sqrt{\rho_{x_A}} \nonumber\\
  \xi'_{x_B^n|x_A^n}&=\Pi^\delta_{C^n|x_A^n}\Pi^\delta_{\hat\rho^n|x_B^n,x_A^n}\hat \rho^n_{x_B^n|x_A^n} \Pi^\delta_{\hat\rho^n|x_B^n,x_A^n}\Pi^\delta_{C^n|x_A^n} \label{eq:xiprime} \\
  \xi'_{x_A^n} &= \sum_{x_B^n}\tilde p_{x_B|x_A}(x_B^n|x_A^n)\xi'_{x_B^n|x_A^n}
  \label{eq:xi}
\end{align}
Here $\Pi^\delta_{\hat\rho^n|x_B^n,x_A^n}$ is
the conditional subspace projector for
the ensemble $\{\Tr\{\Lambda_{x_B|x_A}\rho\},\hat \rho^A_{x_B|x_A}\}$ and
\begin{align}
  \tilde p_{x_B|x_A}(x_B^n|x_A^n)=\begin{cases}
  	\frac 1 {S(x_A^n)} p_{x_B|x_A}(x_B^n|x_A^n) & x_B^n\in T_\delta^{X_B^n|x_A^n} \\
  	0& \text{otherwise}
  \end{cases} \label{eq:pwpruned}
\end{align}
with $S(x_A^n)=\sum_{x_B^n\in T_\delta^{X_B^n|x_A^n}}p_{x_B|x_A}(x_B^n|x_A^n)$.
Let $\Pi$ denote the projector onto the subspace
spanned by the eigenvectors of $\xi'_{x_A}$ with
eigenvalues larger than 
$\epsilon 2^{-n(H(\rho|X_A)+\delta)}=\epsilon 2^{-n(H(R|X_A)+\delta)}$, where the equality is due to the fact that $R$
purifies $C$ for each $x_A$, and
define
\begin{align}
  \Omega_{x_A^n} &= \Pi \xi'_{x_A^n} \Pi \nonumber\\
  \xi_{x_B^n|x_A^n} &= \Pi  \xi'_{x_B^n|x_A^n} \Pi
  \label{eq:Omega}
\end{align}

The shared randomness is generated through two independent
random variables. Call the outcomes $m_A$ and $m_B$. $m_A$ is
shared between Charlie and Alice and potentially Bob. $m_B$ is
shared between Charlie and Bob.

The Bob measurements are generated in one of two ways
\begin{enumerate}
  \item For each outcome $m_B$ of the common randomness, generate
    $s_B'$ sequences $x^n_B(j_B',m_B)$ from the distribution $\tilde p_{x_B}(x_B^n)$. For each $x_A^n\in T_\delta^{X_A^n}$ let $\mathcal{T}(x_A^n)$ be the
    first $s_B$ sequences $x^n_B(j_B',m_B)$ that are
    jointly typical with $x_A^n$; we denote this by
    $x^n_B(j_B,m_B)$. Let $E_c$ be the event that there is
    not at least $s_B$ jointly typical sequences.
  \item For each $x_A^n\in T_\delta^{X_A^n}$ and
each outcome $m_B$ generate $s_B$ sequences $x_B^n(j_B,m_B)$ according
to $\tilde p_{x_B|x_A}(x_B^n|x_A^n)$.
\end{enumerate}

We define the measurement operators
\begin{align}
  \Gamma^{(m_B)}_{x_A^n,j_B} &= \frac{S(x^n_A)}{(1+\epsilon)s_BM_B}\sqrt{\rho_{x_A^n}}^{-1}
  \xi_{x_B^n(j_B,m_B)|x_A^n}\sqrt{\rho_{x_A^n}}^{-1} \nonumber\\
  \Gamma^{(m_A,m_B)}_{j_A,j_B} &= \frac{S(x_A^n(j_A,m_A))}{(1+\epsilon)s_BM_B}\sqrt{\rho_{x_A^n(j_A,m_A)}}^{-1} \nonumber\\
  &\times \xi_{x_B^n(j_B,m_B)|x_A^n(j_A,m_A)}\sqrt{\rho_{x_A^n(j_A,m_A)}}^{-1}
\end{align}
Notice that the operator-valued random variables 
$\xi_{X_B^n(j_B,m_B)|x_A^n}$ are iid. This is clear in case 2,
but also true in case 1, conditioned on $E_c^c$. In fact,
the distribution of $\xi_{X_B^n(j_B,m_B)|x_A^n}$ is exactly
the same as in case 2.

\subsection{Proof part I}
We will show that the set $\Gamma^{(m_A, m_B)}_{j_A}=\{\Gamma^{(m_A,m_B)}_{j_A,j_B}\}_{j_B\in\mathcal{S}}$, where $\mathcal{S}=[s_B]$, is a sub-POVM with high probability for
all $j_A$. If that is not the case, or if in case 1 the event
$E_c$ happens, we put $\Gamma^{(m_B)}_{j_A}=\{I\}$.

We calculate
\begin{align*}
 \MoveEqLeft \sqrt{\rho_{x_A^n(j_A,m_A)}}\sum_{j_B=1}^{s_B}\Gamma^{(m_A,m_B)}_{j_A,j_B}\sqrt{\rho_{x_A^n(j_A,m_A)}} \nonumber\\
  &= \frac{S(x_A^n(j_A,m_A))}{(1+\epsilon)}
  \left(\frac 1 {s_B}\sum_{j_B=1}^{s_B}\xi_{X_B^n(j_B,m_B)|x_A^n(j_A,m_A)}\right)
\end{align*}
where the $X_B^n(j_B,m_B)$ are iid according to $\tilde p_{x_B|x_A}(x_B^n|x_A^n)$. For simplicity of notation,
we denote $x_A^n(j_A,m_A)$ simply by $x_A^n$.
We notice that
\begin{align}
	\MoveEqLeft S(x_A^n)E[\xi_{X_B^n(j_B,m_B)|x_A^n}] \nonumber\\
	&= S(x_A^n)\sum_{x_B^n}\tilde p_{x_B|x_A}(x_B^n|x_A^n)
	\xi_{x_B^n|x_A^n} = S(x_A^n)\Omega_{x_A^n} \nonumber\\
	&=\sum_{x_B^n\in T_\delta^{X_B^n|x_A^n}} p_{x_B|x_A}(x_B^n|x_A^n)\Pi\Pi^\delta_{C^n|x_A^n}\Pi^\delta_{\hat\rho^n|x_B^n,x_A^n}\nonumber\\&\qquad \times \hat \rho^n_{x_B^n|x_A^n} \Pi^\delta_{\hat\rho^n|x_B^n,x_A^n}\Pi^\delta_{C^n|x_A^n}\Pi \nonumber\\
	&\leq\sum_{x_B^n\in T_\delta^{X_B^n|x_A^n}} p_{x_B|x_A}(x_B^n|x_A^n)\Pi\Pi^\delta_{C^n|x_A^n}\hat \rho^n_{x_B^n|x_A^n} \Pi^\delta_{C^n|x_A^n}\Pi \nonumber\\
	&\leq \Pi\Pi^\delta_{C^n|x_A^n}\sum_{x_B^n} p_{x_B|x_A}(x_B^n|x_A^n)\hat \rho^n_{x_B^n|x_A^n} \Pi^\delta_{C^n|x_A^n}\Pi \nonumber\\
	&=\Pi\Pi^\delta_{C^n|x_A^n} \rho^n_{x_A^n} \Pi^\delta_{C^n|x_A^n}\Pi \leq \rho^n_{x_A^n} \label{eq:SOomegaw}
\end{align}
Let $E_m$ be the event that
\begin{align*}
\frac 1{s_B}\sum_{j_B=1}^{s_B}\beta\xi_{X_B^n(j_B,m_B)|x_A^n}\leq \beta\Omega_{x_A^n} (1+\epsilon)
\end{align*}
for some scaling factor $\beta$. By (\ref{eq:SOomegaw})
this event is equivalent to $\sum_{j_B=1}^{s_B}\Gamma^{(m_A,m_B)}_{j_A,j_B}\leq I$, i.e., that $\Gamma^{(m_A, m_B)}_{j_A}$ is 
a sub-POVM. We will show that $E_m$ happens
with high probability using the operator Chernoff bound
\cite[Lemma 17.3.1]{WildeBook}. We notice that by (\ref{eq:SOomegaw}) and the definition
of $\Pi$, $E[\xi_{X_B^n(j_B,m_B)|x_A^n}]=\beta\Omega_{x_A^n}
\geq \beta\epsilon 2^{-n(H(R|X_A)+\delta'')}\Pi$. Furthermore,
\begin{align*}
\beta\xi_{x_B^n|x_A^n}&=\beta\Pi\Pi^\delta_{C^n|x_A^n}\Pi^\delta_{\hat\rho^n|x_B^n,x_A^n}\hat \rho^n_{x_B^n|x_A^n} \Pi^\delta_{\hat\rho^n|x_B^n,x_A^n}\Pi^\delta_{C^n|x_A^n}\Pi \nonumber\\
  &\leq \beta\Pi\Pi^\delta_{C^n|x_A^n}\Pi^\delta_{\hat\rho^n|x_B^n,x_A^n} 2^{-n(H(R|X_A,X_B)-\delta'')} \Pi^\delta_{C^n|x_A^n}\Pi \nonumber\\
  &\leq \Pi 
\end{align*}
when $\beta= 2^{n(H(R|X_A,X_B)-\delta'')}$. The first inequality
follows from properties of conditional quantum typicality
\cite{WildeBook}. Then by the operator Chernoff bound 
\begin{align*}
P(E_m^c) &=
P\left(\frac 1{s_B}\sum_{j_B=1}^{s_B}\beta\xi_{X_B^n(j_B,m_B)|x_A^n}> \beta\Omega_{x_A^n} (1+\epsilon)\right)\nonumber\\
&\leq 2\text{rank}(\Pi)
  \exp\left(-\frac{s_B\epsilon^2\beta\epsilon 2^{-n(H(R|X_A)+\delta)}}{4\ln 2}\right) \nonumber\\
&\leq 2
  \exp\left(-\frac{s_B\epsilon^32^{n(H(R|X_A,X_B)-\delta'')} 2^{-n(H(R|X_A)+\delta'')}}{4\ln 2}\right.\nonumber\\&\left.
  \vphantom{\frac{s\epsilon^32^{n(H(R|W_A)-\delta'')} 2^{-n(H(R|X_A)+\delta'')}}{4\ln 2}}+n(H(R|X_A)+\delta'')\ln 2\right)
\end{align*}
Then
with 
\[
  s_B=2^{n(I(X_B;R|X_A)+3\delta'')} 
\]
the error probability
goes to zero. 

In case 1 we also need to bound $P(E_c)$, the probability that
there is not at least $s_B$ jointly typical sequences. The
number $N$ of jointly typical sequences is binomial with $s_B'$ trials and
success probability about $2^{-nI(X_A;X_B)}$. We can then
bound
\begin{align*}
  P(E_c) &= P(N\leq s_B-1) \nonumber\\
  &\leq \exp\left(-2s_B'\left(2^{-n(I(X_A;X_B)+\delta)}-\frac{s_B-1}{s_B'}\right)\right)
\end{align*}
which converges exponentialy to zero if
\begin{align*}
  s_B'&\geq 2^{n(I(X_A;X_B)+\delta)}s_B \nonumber\\
  &= 2^{n(I(X_A;X_B)+I(X_B;R|X_A)+\delta+3\delta'')}
\end{align*}

The total probability of error then is 
\begin{align*}
  P\left(\bigcup_m E_m^c\right)&\leq \sum_mP(E_m^c)\nonumber\\
  &\leq 2M_B
  \exp\left(-\frac{\epsilon^3 2^{n\delta} }{4\ln 2}+n(H(R|X_A)+\delta)\ln 2\right)
\end{align*}
So, as long as $M_B\leq O(e^n)$, the total error probability
converges to zero.

\subsection{Proof part II: communication}
In case 1, Charlie transmits the index $j_B'$ to Bob. There are $s_B'$ such
indices, so the rate needs to satisfy
\begin{align*}
  R> s_B'> I(X_A;X_B)+I(X_B;R|X_A) = I(X_B;R,X_A)
\end{align*}
In case 1, Charlie transmit the index $j_B$ to Bob. However,
Bob also needs to know the index $j_A$ to decode,
and the rate therefore has to satisfy
\begin{align*}
  R> I(X_A;R)+I(X_B;R|X_A) = I(X_B,X_A;R)
\end{align*}

\subsection{Proof part III: Faithful simulation}
We will prove that we have faithful measurement simulation. 
Let $\tilde\Lambda_{x_A^n}$ be the approximate measurement of $x_A^n$.
Define
\begin{align*}
  \Lambda_{x_B^n}&=\sum_{x_A^n}\sqrt{\Lambda_{x_A^n}}\Lambda_{x_B^n|x_A^n}\sqrt{\Lambda_{x_A^n}}^\dag \nonumber\\
  \tilde\Lambda_{x_B^n}'&=\sum_{x_A^n\in T_\delta^{X_A^n}}\sum_{m_B}\sqrt{\Lambda_{x_A^n}}\sum_{j_B:x_B^n(j_B,m_B)=x^n_B}
  \Gamma^{(m_B)}_{x_A^n,j_B}\sqrt{\Lambda_{x_A^n}}^\dag \nonumber\\
\tilde\Lambda_{x_B^n}&=\sum_{j_A}\sum_{m_A}\sqrt{\Upsilon_{j_A}
^{(m_A)}}\sum_{m_B}\sum_{j_B:x_B^n(j_B,m_B)=x^n_B}\!\!\!\!\!\!\!\!\!\!\!\!\!
  \Gamma^{(m_A,m_B)}_{j_A,j_B}\sqrt{\Upsilon_{j_A}^{(m_A)}}^\dag \nonumber\\
  &= \sum_{x_A^n\in T_\delta^{X_A^n}}\sqrt{\tilde\Lambda_{x_A^n}}\sum_{m_B}\sum_{j_B:x_B^n(j_B,m_B)=x^n_B}
  \Gamma^{(m_B)}_{x_A^n,j_B}\sqrt{\tilde\Lambda_{x_A^n}}^\dag
\end{align*}
We need to bound
\begin{align}
  d=&\sum_{x^n_B}\left\|\sqrt{\rho^n}(\Lambda_{x^n_B}-\tilde\Lambda_{x^n_B})\sqrt{\rho^n}\right\|_1 \nonumber\\
  &\leq \epsilon + \sum_{x^n_B\in T_\delta^{x_B^n}}\left\|\sqrt{\rho^n}(\Lambda_{x^n_B}-\tilde\Lambda_{x^n_B})\sqrt{\rho^n}\right\|_1 \nonumber\\
  &\leq \epsilon + \sum_{x^n_B\in T_\delta^{x_B^n}}\left\|\sqrt{\rho^n}(\Lambda_{x^n_B}-\tilde\Lambda_{x^n_B}')\sqrt{\rho^n}\right\|_1 \nonumber\\
  &+\sum_{x^n_B\in T_\delta^{x_B^n}}\left\|\sqrt{\rho^n}(\tilde\Lambda_{x^n_B}-\tilde\Lambda_{x^n_B}')\sqrt{\rho^n}\right\|_1 \nonumber\\
  &=\epsilon+
  \sum_{x^n_B\in T_\delta^{x_B^n}}\left\|\sum_{x^n_A\in T_\delta^{x_A^n}}\tilde p(x_A^n)\sqrt{\rho_{x_A^n}}\right.\nonumber\\
   &\left.\left(\Lambda_{x_B^n|x_A^n}-\sum_{m_B}\sum_{j_B:x_B^n(j_B,m_B)=x^n_B}
  \Gamma^{(m_B)}_{x_A^n,j_B}\right)\sqrt{\rho_{x_A^n}}\right\|_1\nonumber\\
  &+\sum_{x^n_B\in T_\delta^{x_B^n}}\left\|\sum_{x^n_A\in T_\delta^{x_A^n}}\tilde p(x_A^n)\sqrt{\rho_{x_A^n}}\right.\nonumber\\ &\qquad\qquad\left.\sum_{m_B}\sum_{j_B:x_B^n(j_B,m_B)=x^n_B}
  \Gamma^{(m_B)}_{x_A^n,j_B}\sqrt{\rho_{x_A^n}}\right.\nonumber\\
  &\!\!\!\!\!\!\!\!\left.-\sum_{x^n_A\in T_\delta^{x_A^n}}\hat p(x_A^n)\sqrt{\tilde\rho_{x_A^n}}\sum_{m_B}\sum_{j_B:x_B^n(j_B,m_B)=x^n_B}
  \Gamma^{(m_B)}_{x_A^n,j_B}\sqrt{\tilde\rho_{x_A^n}}\right\|_1
  \label{eq:dbound}
\end{align}
Where we have used that the typical set has probability greater than $1-\epsilon$ and
the triangle inqeuality. We will first bound the last term. Let
\begin{align*}
  \tilde\Gamma_{x_n^A,x_n^B} &= \sum_m\sum_{j_B:x_B^n(j_B,m)=x^n_B}
  \Gamma^{(m)}_{x_A^n,j_B} \nonumber\\
  \mathcal{M}_{x_A^n,x_B^n}(\sigma) 
  &= \sum_{x_B^n}\Tr\{\tilde\Gamma_{x_n^A,x_n^B}\sigma\}
   |x_B^n\rangle\langle x_B^n|
\end{align*}
Also let $\phi_{\rho_{x_A^n}}$ and $\phi_{\tilde\rho_{x_A^n}}$
be purifications of $\rho_{x_A^n}$ and $\tilde\rho_{x_A^n}$
respectively.
Then
\vspace{-0.1in}
\begin{align}
  d_3 &= \sum_{x^n_B\in T_\delta^{x_B^n}}\left\|\sum_{x^n_A\in T_\delta^{x_A^n}}\tilde p(x_A^n)\sqrt{\rho_{x_A^n}}\tilde\Gamma_{x_n^A,x_n^B}\sqrt{\rho_{x_A^n}}\right.\nonumber\\
  &\left.-\sum_{x^n_A\in T_\delta^{x_A^n}}\hat p(x_A^n)\sqrt{\tilde\rho_{x_A^n}}\tilde\Gamma_{x_n^A,x_n^B}\sqrt{\tilde\rho_{x_A^n}}\right\|_1 \nonumber
\end{align}
\begin{align}
  &= \left\|\sum_{x^n_A\in T_\delta^{x_A^n}}\tilde p(x_A^n)
  (I^R\otimes \mathcal{M}_{x_A^n,x_B^n})(\phi_{\rho_{x_A^n}})\right.\nonumber\\
  &\left.-\sum_{x^n_A\in T_\delta^{x_A^n}}\hat p(x_A^n)(I^R\otimes \mathcal{M}_{x_A^n,x_B^n})(\phi_{\tilde\rho_{x_A^n}})\right\|_1 \nonumber\\
  &\leq \sum_{x^n_A\in T_\delta^{x_A^n}}
  |\tilde p(x_A^n)-\hat p(x_A^n)| \nonumber\\
  &+\sum_{x^n_A\in T_\delta^{x_A^n}}
  \tilde p(x_A^n)\left\|
  (I^R\otimes \mathcal{M}_{x_A^n,x_B^n})(\phi_{\rho_{x_A^n}})\right.\nonumber\\
  &\left.-(I^R\otimes \mathcal{M}_{x_A^n,x_B^n})(\phi_{\tilde\rho_{x_A^n}})\right\|_1 \nonumber\\
  &\leq \sum_{x^n_A\in T_\delta^{x_A^n}}
  |\tilde p(x_A^n)-\hat p(x_A^n)| +\sum_{x^n_A\in T_\delta^{x_A^n}}
  \tilde p(x_A^n)\left\|
  \phi_{\rho_{x_A^n}}-\phi_{\tilde\rho_{x_A^n}}\right\|_1
  \label{eq:d3}
\end{align}
The first equality follows from a slightly modified version of
\cite[Lemma 4]{WildeAl2012} and the first inequality from \cite[Exercise 9.1.11]{WildeBook}.
The second inequality is because $\mathcal{M}_{x_A^n,x_B^n}$ is a contraction,
$
  \|\mathcal{M}_{x_A^n,x_B^n}(\sigma)\|_1\leq \|\sigma\|_1
$.
Now
\begin{align*}
  \|\phi_{\rho_{x_A^n}}-\phi_{\tilde\rho_{x_A^n}}\|_1
  &= \sqrt{1-|\langle\phi_{\rho_{x_A^n}}|\phi_{\tilde\rho_{x_A^n}}\rangle|^2} = \sqrt{1-F(\rho_{x_A^n},\tilde\rho_{x_A^n})}
\end{align*}
where $F$ is fidelity and we use the canonical purification. Since
$
  F(\rho_{x_A^n},\tilde\rho_{x_A^n})\geq 
  \left(1-\frac 1 2\|\rho_{x_A^n}-\tilde\rho_{x_A^n}\|_1\right)^2
$,
\begin{align*}
\MoveEqLeft  \sum_{x_A^n}\tilde p(x_A^n)\|\phi_{\rho_{x_A^n}}-\phi_{\tilde\rho_{x_A^n}}\|_1 \nonumber\\
  &\leq \sqrt{1-\left(1-\frac 1 2\sum_{x_A^n}\tilde p(x_A^n)\|\rho_{x_A^n},\tilde\rho_{x_A^n}\|_1\right)^2} \nonumber\\
  &\leq \sqrt{1-\left(1-\epsilon\right)^2}=\epsilon'
\end{align*}
because $\sqrt{1-\left(1-\frac 1 2 x\right)^2}$ is a concave function and
(\ref{eq:rxA2}).
As the first term in (\ref{eq:d3}) is bounded by $\epsilon$ in (\ref{eq:pA}), then
$d_3\leq \epsilon+\epsilon'$.

\todo{SEVERAL THING TO VERIFY ABOVE}

We next bound the second term in (\ref{eq:dbound}). 
\begin{align}
  d_2 &= \sum_{x^n_B\in T_\delta^{x_B^n}}\left\|\sum_{x^n_A\in T_\delta^{x_A^n}}\tilde p(x_A^n)\sqrt{\rho_{x_A^n}}\left(\Lambda_{x_B^n|x_A^n}\right.\right.\nonumber\\ &\left.\left.-\sum_{m_B}\sum_{j_B:x_B^n(j_B,m_B)=x^n_B}
  \Gamma^{(m_B)}_{x_A^n,j_B}\right)\sqrt{\rho_{x_A^n}}\right\|_1 \nonumber
\end{align}
\begin{align}
  &\leq \sum_{x^n_B\in T_\delta^{x_B^n}}\sum_{x^n_A\in T_\delta^{x_A^n}}\tilde p(x_A^n)\left\| p(x_B^n|x_A^n)\hat\rho_{x_B^n|x_A^n}\right.\nonumber\\&\left.-\frac{S(x^n_A)c(x_B^n|x_A^n)}{(1+\epsilon)s_BM_B}\xi_{x_B^n|x_A^n}\right\|_1 \nonumber\\
  &\leq \sum_{x^n_B\in T_\delta^{x_B^n}}\sum_{x^n_A\in T_\delta^{x_A^n}}\tilde p(x_A^n) p(x_B^n|x_A^n)\left\|\hat\rho_{x_B^n|x_A^n}-\xi_{x_B^n|x_A^n}'\right\|_1 \nonumber\\
  &+\sum_{x^n_B\in T_\delta^{x_B^n}}\sum_{x^n_A\in T_\delta^{x_A^n}}\tilde p(x_A^n) p(x_B^n|x_A^n)\left\|\xi_{x_B^n|x_A^n}'-\xi_{x_B^n|x_A^n}\right\|_1 \nonumber\\
  &+ \sum_{x^n_B\in T_\delta^{x_B^n}}\sum_{x^n_A\in T_\delta^{x_A^n}}\tilde p(x_A^n)\left\| p(x_B^n|x_A^n)\xi_{x_B^n|x_A^n}\right.\nonumber\\&\left.-\frac{S(x^n_A)c(x_B^n|x_A^n)}{(1+\epsilon)s_BM_B}\xi_{x_B^n|x_A^n}\right\|_1 \nonumber\\  
  &\leq \sum_{x^n_B\in T_\delta^{x_B^n}}\sum_{x^n_A\in T_\delta^{x_A^n}}\tilde p(x_A^n) p(x_B^n|x_A^n)\left\|\hat\rho_{x_B^n|x_A^n}-\xi_{x_B^n|x_A^n}'\right\|_1 \nonumber\\
  &+\sum_{x^n_B\in T_\delta^{x_B^n}}\sum_{x^n_A\in T_\delta^{x_A^n}}\tilde p(x_A^n) p(x_B^n|x_A^n)\left\|\xi_{x_B^n|x_A^n}'-\xi_{x_B^n|x_A^n}\right\|_1 \nonumber\\
  &+ \sum_{x^n_B\in T_\delta^{x_B^n}}\sum_{x^n_A\in T_\delta^{x_A^n}}\tilde p(x_A^n)\left| \frac{p(x_B^n|x_A^n)}{S(x^n_A)}-\frac{c(x_B^n|x_A^n)}{(1+\epsilon)s_BM_B}\right| \label{eq:d2}
\end{align}
We  bound the third term in (\ref{eq:d2}). We use the operator Chernoff bound
\cite[Lemma 17.3.1]{WildeBook}. For every $x_A^n\in T_\delta^{X_A^n}$ let $P_{x_A^n}$ be
the diagonal matrix with $\tilde p(x_B^n|x_A^n)$ on
the diagonal for all $x_B^n\in T_\delta^{X_B^n|x_A^n}$, and let
$C_{x_A^n}$ be the same for  the empirical frequencies $\frac{c(x_B^n|x_A^n)}{s_BM_B}$. We have $E[C_{x_A^n}]=P_{x_A^n}$ and
$P_{x_A^n}\geq 2^{-n(H(X_B|X_A)+\delta'')}I$. Let $E_0$ be the event
that for all $x_A^n\in T_\delta^{X_A^n}$
\begin{align*}
  (1-\epsilon)P_{x_A^n}\leq C_{x_A^n}\leq (1+\epsilon)P_{x_A^n}
\end{align*}
The operator Chernoff bound and the union bound then gives
\begin{align*}
  P(E_0^c) &\leq 2\ddot \cdot 2^{n(H(X_A)+\delta)} \cdot 2^{-n(H(X_B|X_A)+\delta'')} \nonumber\\
  &\times\exp\left(-\frac{M_Bs_B\epsilon 2^{-n(H(X_B|X_A)+\delta'')}}{4\ln 2}\right)
\end{align*}
Thus, if 
$
M_Bs_B\geq 2^{-n(H(X_B|X_A)+2\delta'')}, 
$
this probability
converges to zero. Then
conditioned on $E_0^c$:
\begin{align*}
\MoveEqLeft
  \sum_{x_B^n\in T_\delta^{X_B^n|x_A^n}}
  \left| \frac{p(x_B^n|x_A^n)}{S(x^n_A)}-\frac{c(x_B^n|x_A^n)}{(1+\epsilon)s_BM_B}\right|\nonumber\\
  & = \left\|\frac 1{1+\epsilon}C_{x_A^n}-P_{x_A^n}\right\|_1 \nonumber\\
  &\leq \frac 1{1+\epsilon}(\|\epsilon P_{x_A^n}\|_1+\|C_{x_A^n}-P_{x_A^n}\|_1) \leq \frac{2\epsilon}{1+\epsilon}
\end{align*}
which bounds the last term in (\ref{eq:d2}). The first two terms in (\ref{eq:d2})
are bounded just as in \cite[(35)]{WildeAl2012}
using the modified definition (\ref{eq:Omega}).

\bibliographystyle{IEEEtran}
\bibliography{Coop06,ahmref2,Coop03,BigData,Quantum}

\end{document}